\documentclass[12pt,a4paper]{article}
\usepackage{amsmath,amssymb,hyperref,bm,bbm}
\usepackage{braket,color,ascmac,mathrsfs}
\usepackage[utf8]{inputenc}
\usepackage{graphicx}
\usepackage{cite}
\numberwithin{equation}{section}
\allowdisplaybreaks[3]

\newcommand{\de}{\partial}
\newcommand{\be}{\begin{equation}}
\newcommand{\ba}{\begin{eqnarray}}
\newcommand{\ea}{\end{eqnarray}}
\newcommand{\ee}{\end{equation}}

\newcommand{\s}{\sqrt}

\newcommand{\ap}{\alpha}

\newcommand{\ddd}{\cdot\cdot\cdot}
\newcommand{\no}{\nonumber \\}
\newcommand{\la}{\langle}
\newcommand{\lb}{\rangle}
\newcommand{\bea}{\begin{eqnarray}}
\newcommand{\eea}{\end{eqnarray}}
\newcommand{\bes}{\begin{equation*}}
\newcommand{\beas}{\begin{eqnarray*}}
\newcommand{\eeas}{\end{eqnarray*}}
\newcommand{\bas}{\begin{array*}}
\newcommand{\eas}{\end{array*}}
\newcommand{\ees}{\end{equation*}}

\newcommand{\ep}{\epsilon}

\begin{document}
\begin{titlepage}
		
\renewcommand{\thefootnote}{\fnsymbol{footnote}}
\begin{flushright}
\begin{tabular}{l}
YITP-20-143
\\
IPMU20-0113
\\
\end{tabular}
\end{flushright}
		
\vfill
\begin{center}
			
			
\noindent{\Large \textbf{Chern-Simons Gravity Dual of BCFT}}

\vspace{1.5cm}

\noindent{Tadashi Takayanagi$^{a,b,c}$\footnote{E-mail: takayana@yukawa.kyoto-u.ac.jp} and Takahiro Uetoko$^a$\footnote{E-mail: takahiro.uetoko@yukawa.kyoto-u.ac.jp}}
\bigskip

\vskip .6 truecm
\centerline{\it $^a$Center for Gravitational Physics, Yukawa Institute for Theoretical Physics,}
\centerline{\it  Kyoto University, Kyoto 606-8502, Japan}
\medskip
\centerline{\it $^b$Inamori Research Institute for Science, Kyoto 600-8411, Japan}
\medskip
\centerline{\it $^c$Kavli Institute for the Physics and Mathematics of the Universe (WPI),}
\centerline{\it University of Tokyo, Chiba 277-8582, Japan}
			
\end{center}
		
\vfill
\vskip 0.5 truecm

\begin{abstract}

In this paper we provide a Chern-Simons gravity dual of a two dimensional conformal field theory on a manifold with boundaries, so called boundary conformal field theory (BCFT). 
We determine the correct boundary action on the end of the world brane 
in the Chern-Simons gauge theory. This reproduces known results of the AdS/BCFT for the Einstein gravity. We also give a prescription of calculating holographic entanglement entropy by employing Wilson lines which extend from the AdS boundary to the end of the world brane. We also discuss a higher spin extension of our formulation.

\end{abstract}
\vfill
\vskip 0.5 truecm
		
\setcounter{footnote}{0}
\renewcommand{\thefootnote}{\arabic{footnote}}
\end{titlepage}
	
\newpage
	
\tableofcontents

	
\section{Introduction}

The AdS/CFT correspondence \cite{Maldacena:1997re} has enabled us to connect the dynamics of a $d$ dimensional conformal field theory (CFT) 
to that of gravity on a $d+1$ dimensional Anti de-Sitter space (AdS)
in a surprising way. We can extend the AdS/CFT to the case where a CFT is defined on a manifold with boundaries, called BCFT (boundary conformal field theory) \cite{Cardy:2004hm}, by imposing a Neumann boundary condition on a surface in the bulk (called the end of the world brane) \cite{Karch:2000gx,Takayanagi:2011zk,Fujita:2011fp}. 
This formulation has been called the AdS/BCFT construction and has had 
many applications such as renormalization group flow 
\cite{Gutperle:2012hy,Estes:2014hka,Kobayashi:2018lil,Sato:2020upl},  
condensed matter models \cite{Fujita:2012fp,Erdmenger:2015xpq}, non-equilibrium physics including quantum entanglement \cite{Ugajin:2013xxa,Seminara:2017hhh,Seminara:2018pmr,Hikida:2018khg,Shimaji:2018czt,Caputa:2019avh,Mezei:2019zyt}, and computational complexity \cite{Chapman:2018bqj,Sato:2019kik,Braccia:2019xxi}. Refer also to \cite{Chiodaroli:2011nr,Chiodaroli:2012vc} for string theory embeddings of gravity duals of BCFTs and 
to \cite{Karch:2020iit,Bachas:2020yxv,Simidzija:2020ukv,Ooguri:2020sua} for 
implications to novel string theory backgrounds.

In the holographic entanglement entropy \cite{Ryu:2006bv,Ryu:2006ef,Hubeny:2007xt} for AdS/BCFT, the minimal surface, whose area computes the holographic entanglement entropy, can terminate on the end of the world brane \cite{Takayanagi:2011zk,Fujita:2011fp}. This leads to the structure so called the Island \cite{Penington:2019npb,Almheiri:2019psf,Almheiri:2019hni}, which has been successfully applied to studies of black hole information \cite{Rozali:2019day,Almheiri:2019psy,Chen:2020uac,Bousso:2020kmy,Geng:2020qvw}. Moreover, limits of AdS/BCFT lead to a series of codimension two holography \cite{Bousso:2020kmy,Akal:2020wfl,Miao:2020oey}
(see also \cite{Takayanagi:2019tvn}).

For $d=2$ dimensional BCFTs,  a wide variety of AdS/BCFT setups can be studied analytically, while in higher dimensions, analytical examples are highly limited \cite{Nozaki:2012qd,Chu:2017aab}. It is natural to expect that this special feature of $d=2$ stems from the mathematically beautiful fact that we can formulate three dimensional gravity in terms of SL$(2,R)$ Chern-Simons gauge theory \cite{Achucarro:1987vz,Witten:1988hc}. This Chern-Simons formulation has an important advantage that we can generalize the Einstein gravity to higher spin gravity by replacing the gauge group with SL$(n,R)$ \cite{Blencowe:1988gj,Campoleoni:2010zq,Henneaux:2010xg}, which is a door to studying quantum gravity.

Motivated by this, we present a formulation of AdS/BCFT in the Chern-Simons gauge theory description of three dimensional gravity in this paper. For this we will introduce the end of the world-brane in the SL$(2,R)$ Chern-Simons theory and identity its action with correct boundary terms which reproduce the expected results of AdS/BCFT. Next we will consider the holographic entanglement entropy, which provides us with a useful probe of bulk geometry, expressed in terms of quantum information of the dual CFT. In the Chern-Simons description of gravity, including higher spin gravity, the holographic entanglement entropy is remarkably calculated as an expectation value of a certain Wilson loop
\cite{Ammon:2013hba,deBoer:2013vca}. See also e.g. \cite{Datta:2014ska,Bagchi:2014iea,deBoer:2014sna,Castro:2015csg,Chen:2016uvu,Jiang:2017ecm,Huang:2019nfm} for later progresses and \cite{Fitzpatrick:2016mtp,Besken:2017fsj,Hikida:2017ehf,Hikida:2018dxe,Hikida:2018eih,Besken:2018zro} for related works including quantum corrections. For higher spin black hole entropy refer to e.g. \cite{Gutperle:2011kf,Perez:2012cf,Campoleoni:2012hp,deBoer:2013gz,Kraus:2013esi}.
We will extend this calculation of holographic entanglement entropy in Chern-Simons gravity to the case where we have the end of the world brane in the three dimensional bulk. We will find that we can view the end of the world brane can be regarded as a sort of ''D-brane'' in Chern-Simons gauge theory.

This paper is organized as follows.  In section two we briefly review the AdS/BCFT construction and its holographic entanglement entropy for the Einstein gravity. In section three, we present Chern-Simons gravity  formulation of AdS/BCFT.  In section four, we give the prescription of calculating  holographic entanglement entropy in Chern-Simons gravity in the presence of the end of the world-brane. In section five, we summarize our conclusions and discuss future problems. In appendix A, we present our convention of Lie algebras. In appendix B,  we present the details of rewriting the Einstein gravity in terms of Chern-Simons gauge theory. In appendix C, we give a brief summary of the geodesic length in AdS$_3$ for a convenience.


\section{A Brief Review of AdS/BCFT}

In this section we would like to give a brief summary of a bottom up construction  of gravity duals 
for CFTs on manifolds with boundaries, called the AdS/BCFT \cite{Takayanagi:2011zk,Fujita:2011fp}.
We focus on the $d=2$ case where BCFT is a two dimensional CFT with a boundary and its gravity dual
 is three dimensional. We can generalize our construction here to a case with multiple boundaries
in straightforward way and we will omit this extension below.

\subsection{Construction of AdS/BCFT}

\begin{figure}
  \centering
  \includegraphics[bb=0 0 960 540,width=10cm]{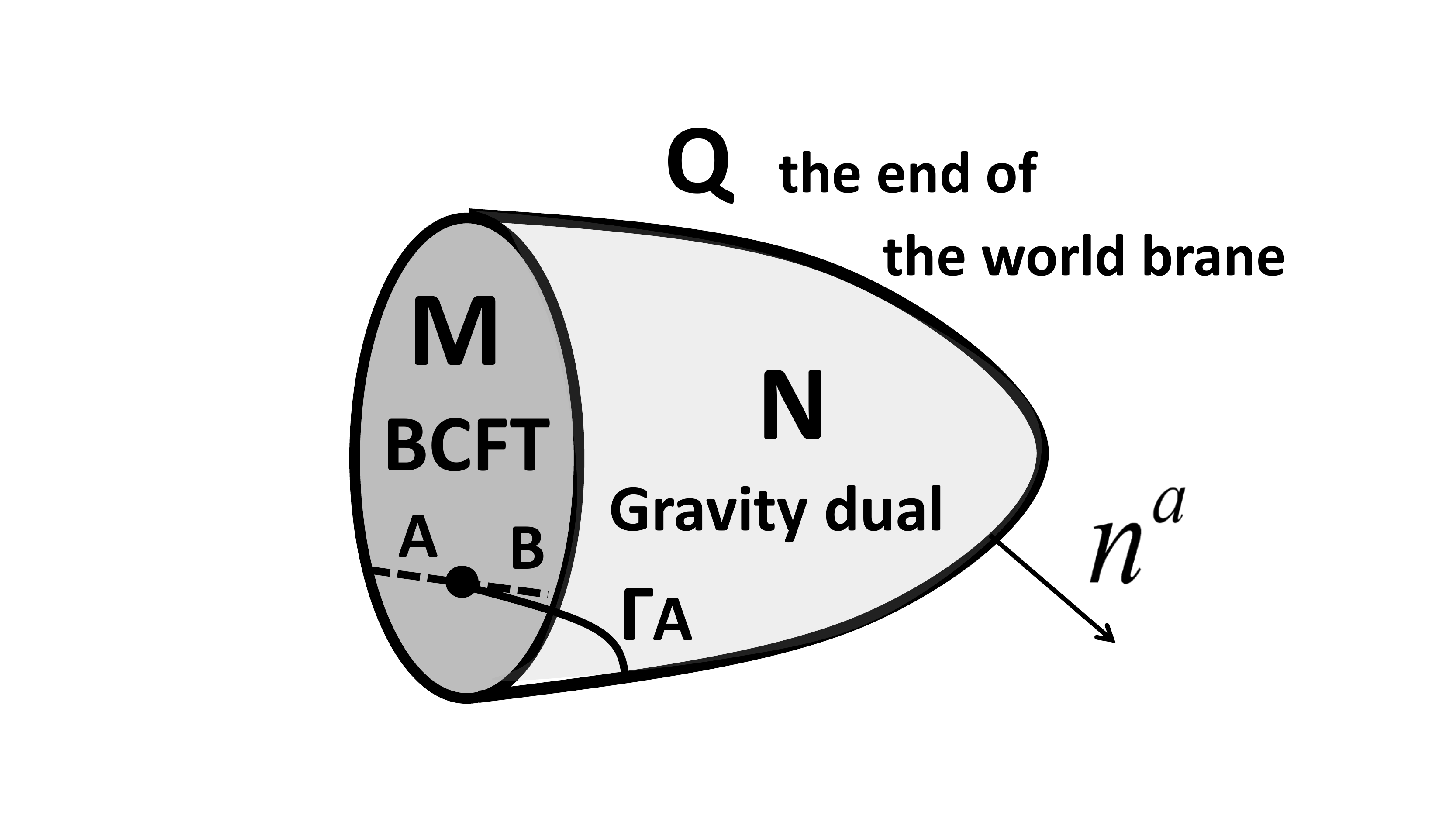}
  \caption{A sketch of AdS/BCFT. We have in mind a $d=2$ example.
A BCFT is defined on the manifold $M$ and its gravity dual is given by the region $N$. The boundary of $M$ is extended into the bulk as the surface $Q$, called the end of the world brane. The dotted line represents a time slice on the boundary $M$, which is divided into two subregions $A$ and $B$. The entanglement entropy $S_A$ in the BCFT 
can be holographically computed as  the geodesic length of $\Gamma_A$ which starts from the boundary 
$\de A$ and ends on a point on the surface $Q$, where we minimize the geodesic length by changing the location of the point on $Q$. }
\label{bcftafig}
\end{figure}

Consider a BCFT on a two dimensional manifold $M$ with a boundary $\de M$. The AdS/BCFT argues that 
its gravity dual is given by a gravity on a three dimensional space $N$ which satisfies 
\ba
\de N=M\cup Q,\ \ \ s.t. \ \ \ \de Q=\de M.
\ea
Refer to Fig.\ref{bcftafig} for an illustration. The manifold $M$ coincides with the asymptotically AdS boundary of $N$. 
The surface $Q$, which surrounds the gravity dual is called the end of the world brane and is a bulk extension of the boundary $\de M$. The metric on $N$ and the shape of $Q$ are determined by the equation of motions 
derived from the bulk gravity action
\ba
I_{\mbox{gravity}}=I_{EH}+I_{GH},
\ea
where $I_{EH}$ is the Einstein-Hilbert action
\begin{align}
	I_{EH}=\frac{1}{16\pi G_N}\int_N \sqrt{-g}(R-2\Lambda)~,
\label{eq:IEH}
\end{align}
and $I_{GH}$ is the Gibbons-Hawking term with the boundary cosmological constant term\footnote{
We can also introduce matter fields localized on the surface $Q$ \cite{Takayanagi:2011zk,Fujita:2011fp}.
However, we only care about the pure gravity degrees of freedom in this paper for the Chern-Simons description and thus we do not turn on such matter fields.}:
\begin{align}
	I_{GH}=\frac{1}{8\pi G_N}\int_{\partial N} \sqrt{-h}(K-T)~.
\label{eq:IGH}
\end{align}
Here $g_{\mu\nu}$ and $h_{ij}$ is the metric of $N$ and $Q$. The tensor $K_{ij}$ is the extrinsic curvature 
of $Q$ and $K=h^{ij}K_{ij}$ is its trace. The extrinsic curvature is defined by the projection on $Q$ of the covariant derivative of the out-going unit vector $n$, normal to  the surface $Q$. The constant $T$ can be interpreted as the tension of the brane $Q$. 

By varying $I_{EH}$ with respect to the bulk metric $g$, we obtain the Einstein equation as usual.
Moreover, by taking the boundary variation with respect to the boundary metric $h$, we obtain
\begin{align}
\delta (I_{EH}+I_{GH})=\frac{1}{16\pi G}\int_{\partial N}\sqrt{-h}(K_{ij}-(K-T) h_{ij})\delta h^{ij}~.
\end{align}
By setting this to be vanishing, we obtain a boundary condition which is either 
 the Dirichlet boundary condition
\begin{align}
	\delta h^{ij}=0, \label{dbc}
\end{align}
or the Neumann one
\begin{align}
	K_{ij}-(K-T) h_{ij}=0~.  \label{nbc}
\end{align}
For the AdS boundary $M$, we impose the Dirichlet boundary condition (\ref{dbc}) as usual. 
On the other hand, we impose the Neumann boundary condition (\ref{nbc}) for the boundary $Q$, i.e. the end of the world brane. In this way, the bulk metric $g$ and the profile of the surface $Q$ is determined by solving the bulk Einstein equation together with  the mentioned boundary conditions on $M$ and $Q$. This is a short summary of AdS/BCFT construction.

\subsection{Holographic Entanglement Entropy in AdS/BCFT}\label{sec:heeb}

The entanglement entropy $S_A$ for a subsystem $A$ is defined from a pure quantum state $|\Psi\lb$ is defined by
\ba
S_A=-\mbox{Tr}[\rho_A\log\rho_A].
\ea
We introduced $\rho_A$, called the reduced density matrix
\ba
\rho_A=\mbox{Tr}_B[|\Psi\lb\la\Psi|],
\ea
where $B$ is the complement of $A$. We can define the entanglement entropy in a BCFT as is so in standard field theories by taking a specific time slice and dividing the time slice into two parts $A$ and $B$.
For example, consider a BCFT on a semi-infinite line (times a time direction) 
which ends on a boundary. If we choose  the subsystem $A$ to be a length $l$ interval which extends from the boundary, then the entanglement entropy $S_A$ is known to be given by the following formula \cite{Calabrese:2004eu}:
\ba
S_A=\frac{c}{6}\log\frac{l}{\ep}+S_{bdy},  \label{bcftee}
\ea
where $c$ is the central charge of the two dimensional CFT  and $\ep$ is the UV cut off (i.e. the lattice spacing). The constant term $S_{bdy}$ coincides with an important quantity called the boundary entropy 
\cite{Affleck:1991tk}, which measures the amount of the degrees of freedom of the boundary.

The holographic dual of the entanglement entropy $S_A$ \cite{Ryu:2006bv,Ryu:2006ef,Hubeny:2007xt} in the AdS/BCFT  construction is computed from  the following formula \cite{Takayanagi:2011zk,Fujita:2011fp}:
\ba
S_A=\mbox{Min}\ \mbox{Ext}\left[\frac{\mbox{Area}(\Gamma_A)}{4G_N}\right],\ \ s.t.\ \  
\de\Gamma_A=\de A\cup \de \Sigma,\ \ \ \Sigma\in Q.  \label{RTB}
\ea
where Min and Ext mean that we first extremize the  area of $\Gamma_A$ and after that we pick up the minimal one among discrete candidates of extremal surface. The crucially new aspect peculiar to the 
AdS/BCFT, is that the extremal surface $\Gamma_A$ can end on the surface $Q$. The region on $Q$
which is surround by $\Gamma_A$ is written as $\Sigma$ in (\ref{RTB}), which is recently called an Island \cite{Penington:2019npb,Almheiri:2019psf,Almheiri:2019hni}. In our $d=2$ case, $\Gamma_A$ is a geodesic which connects $\de A$ and $Q$, where $\de\Sigma$ degenerates into a point on $Q$ as depicted in Fig.\ref{bcftafig}. Therefore to calculate the holographic entanglement entropy we need to extremize with respect to the location of a point on $Q$ on which the geodesic $\Gamma_A$ ends. 
This prescription successfully reproduces the expression 
(\ref{bcftee}) \cite{Takayanagi:2011zk,Fujita:2011fp} and we will reproduce this in our Chern-Simons 
gravity formalism later in this paper.


\section{Chern-Simons Gravity and Boundary Condition}

In this section, we describe a three dimensional Einstein gravity on a manifold with boundaries in terms of 
Chern-Simons (CS) gravity and determine the boundary condition. We assume a negative cosmological 
constant. In terms of the CS valuables with $G=SL(2,R)\times SL(2,R)$ we can rewrite $I_{EH}$ (\ref{eq:IEH}) 
as \cite{Achucarro:1987vz,Witten:1988hc}
\begin{align}
	I_{EH}=\frac{k}{4\pi}\int_N \text{Tr}\left[ A\wedge dA+\frac{2}{3}A^3\right]-\frac{k}{4\pi}\int_N \text{Tr}\left[ \bar{A}\wedge d\bar{A}+\frac{2}{3}\bar{A}^3\right]+\frac{k}{4\pi}\int_{\partial{N}}\text{Tr} \left[A\wedge \bar{A}\right],
\label{eq:ICS1}
\end{align}
where $A$ and $\bar{A}$ can be decomposed into two pairs of connections as
\begin{align}
	A=A^aJ_a=\left(\omega^a+e^a\right)J_a,\quad \bar{A}=\bar{A}^aJ_a=\left(\omega^a-e^a\right)J_a.
\label{eq:Ae}
\end{align}
Here $J_a\in sl(2,R)$ are explicitly given by
\ba
 J_1=\left(
 \begin{array}{cc}
  0 & 0 \\
  -1 & 0 \\
 \end{array} 
\right),\quad J_{0}=\left(
 \begin{array}{cc}
  1/2 & 0 \\
  0 & -1/2 \\
 \end{array} 
\right),\quad J_{-1}=\left(
 \begin{array}{cc}
  0 & 1 \\
  0 & 0 \\
 \end{array} 
\right),
\label{a2base}
\ea
and we also define
\ba
J_\rho=J_0, \ \ \ J_t= \frac{J_1+J_{-1}}{2},\ \ \ J_\phi=\frac{J_1-J_{-1}}{2},
\ea
which obeys
\ba
[J_\rho,J_t]=-J_\phi,\ \ \  [J_t,J_\phi]=-J_\rho,\ \ \ [J_\phi,J_\rho]=J_t.
\ea
We write the three dimensional coordinate $x^\mu$ as $(x^0,x^1,x^2)=(t,\phi,\rho)$.
Note that $k$ is the CS level with $k=\frac{1}{4G_N}$, see also details of this formulation in appendix \ref{app:CSform}.

\subsection{Adding Boundary Term}

The correct CS gravity action with a boundary $Q$ is given by\footnote{Here we omit the boundary terms for the standard AdS boundary $M$.}
\begin{align}
	I_{tot}^{CS}=I_{EH}+I_{GH}
\label{eq:Itot}
\end{align}
with (\ref{eq:IEH}) and (\ref{eq:IGH}). 
$I_{GH}$ is rewritten as\footnote{Refer to an earlier work \cite{Corichi:2015nea} for the second term in (\ref{eq:ICS2}).}\footnote{We can confirm the following equation
\begin{align}
	\int_Q \sqrt{-h} K =-\int_{Q}  P_{ab}e^a\wedge\omega^b-\int_{Q}\epsilon_{abc}e^a\wedge n^b dn^c
\end{align}
with this $n^a$.}
\begin{align}
	I_{GH}=-\frac{k}{2\pi}\int_Q P_{ab}e^a\wedge\omega^b-\frac{k}{2\pi}\int_Q \epsilon_{abc}e^a\wedge n^b dn^c-\frac{k T}{4\pi}\int_Q\epsilon_{abc}n^a e^b\wedge e^c~,
\label{eq:ICS2}
\end{align}
where $\epsilon_{abc}$ is the epsilon tensor such that $\epsilon_{t\phi\rho}=1$ and $n^a$ is a vector with a unit norm $n^an_a=1$. $P_{ab}$ is the projector on $Q$ defined as $P_{ab}=\eta_{ab}-n_an_b$ with $\eta_{ab}=2\text{Tr}[J_aJ_b]$. Also we simplified $\int_Q d\rho\wedge dt$ as $\int_Q$.
Note here that, at first, we treat $n^a$ as a unit vector field on $Q$ and it is regarded as a dynamical field. After we impose the boundary condition we will find that $n^a$ is a normal vector of 
the surface $Q$. 

With (\ref{eq:ICS1}) and (\ref{eq:ICS2}), we find the total variation\footnote{We employed the Stokes' theorem: $\int_M d(A\wedge \delta A)=-\int_Q A\wedge \delta A$.}
\begin{align}
\begin{aligned}
	\delta I^{CS}_{tot}&=-\frac{k}{2\pi}\int_Q \delta e^a \wedge (P_{ab}\omega^b+\epsilon_{abc}n^b dn^c-T\epsilon_{abc}n^b e^c)+\frac{k}{2\pi} \int_Q n_an_b e^a\wedge \delta \omega^b \\
	&\qquad+\frac{k}{4\pi}\int_Q \delta n^a\left[4\epsilon_{abc}e^b dn^c+4n_b e^b\wedge \omega_a-T\epsilon_{abc}e^b\wedge e^c\right],
\label{eq:dI}
\end{aligned}
\end{align}
where we use the relation
\begin{align}
	\epsilon_{abc}n^b (de^c)=(n_be^b)\wedge \omega^a-e^a\wedge (n_b\omega^b)
\end{align}
with the equation of motion 
\begin{align}
	de^a-\epsilon^a_{bc}\omega^b\wedge e^c=0.
\end{align}
This leads to the following (Neumann) boundary conditions on $Q$:
\begin{align}
	&P_{ab}\omega^b+\epsilon_{abc}n^b dn^c-T\epsilon_{abc}n^b e^c=0~, \label{eq:NC1} \\
	&n_a e^a=0~, \label{eq:NC2} \\
	&P_{a'a}(4\epsilon^a_{bc}e^b dn^c+4n_b e^b\wedge \omega^a-T\epsilon^a_{bc}e^b\wedge e^c)=0~. \label{eq:NC3}
\end{align}
The second condition (\ref{eq:NC2}) says that $n^a$ is a normal vector on $Q$ i.e. if we define the normal unit vector on $Q$ by $N_\mu$, then the vector $n^a$ is defined by 
\begin{align}
	n^a=e^a_\mu N^\mu.
\end{align}
Note also that we acted $P_{a'a}$ in the third condition (\ref{eq:NC3}) to project to the direction orthogonal to $n^a$. This is simply because the variation $\delta n^a$ is restricted to $n_a\delta n^a=0$ because $n^an_a=1$. The third condition (\ref{eq:NC3}) is trivially satisfied if we assume  (\ref{eq:NC2})  and remember that $n_a dn^a=0$. Therefore once we choose $n^a$ to be the normal vector on $Q$, we have only to solve the first equation (\ref{eq:NC1}), which is equivalent to (\ref{nbc}) as we show in appendix 
\ref{app:CSform}.

\subsection{Gauge Invariance}

Let us consider the gauge transformation
\ba
\delta A=d\chi+[A,\chi],\ \ \ \ \delta\bar{A}=d\bar{\chi}+[\bar{A},\bar{\chi}]. \label{gtqa}
\ea
If write $\chi=\ap+\beta$ and $\bar{\chi}=\ap-\beta$, then the gauge transformation looks like
\ba
&&\delta e=d\beta+[\omega,\beta]+[e,\ap],\no
&& \delta \omega=d\ap+[\omega,\ap]+[e,\beta].
\ea

First consider the variation of Einstein-Hilbert action $I_{EH}$ i.e, (\ref{eq:ICS1}), which looks like
\ba
\delta I_{EH}
=\frac{k}{4\pi}\int_Q (\bar{\chi}-\chi)\wedge 
[dA+d\bar{A}+A\wedge \bar{A}+\bar{A}\wedge A].
\ea
Therefore we find that this is vanishing if we impose the boundary condition 
\ba
(\chi-\bar{\chi})|_Q=0.
\ea
Thus the part $\ap$ of the gauge transformation is still active at the boundary.

The remaining term of $I_{GH}$ (\ref{eq:ICS2}) is 
\ba
I_{GH}=-\frac{k}{2\pi}\int_Q \ep_{abc}e^a\wedge n^b (dn^c+\ep^c_{pq} \omega^p n^q)-\frac{kT}{4\pi}\int \ep_{abc}n^a e^b \wedge e^c.
 \label{bdyasss}
\ea
It is clear that this term is invariant  under the $\ap$ gauge transformation (i.e. the local $SL(2,R)$ transformation), which looks like
\ba
\delta e=[e,\ap],\ \ \ \delta n=[n,\ap],\ \ \ \delta \omega=d\ap+[\omega,\ap].
\ea 
Indeed, because $[J_a,J_b]=-\ep^c_{ab}J_c$, we find the $\ap$ gauge transformation leads to
\ba
\delta (dn-[\omega,n])=[dn-[\omega,n],\ap].
\ea 
Thus we can regard $dn-[\omega,n]=Dn^aJ_a$ as the covariant derivative \cite{Corichi:2015nea} where
\ba
Dn^a=dn^a+\ep^a_{pq} \omega^p n^q,
\ea
which behaves like a vector under the local $SL(2,R)$ transformation in the same way as $e^a$ and $n^a$.

\subsection{Solutions to Boundary Condition}

In this paper, we focus on examples constructed from the Poincare AdS$_3$
\ba
ds^2=e^{2\rho}(-dt^2+d\phi^2)+d\rho^2.  \label{AdS3m}
\ea
We write the gauge fields
\begin{align}
	A=e^{\rho}J_{1}dx^++J_{0}d\rho,\quad \bar{A}=-e^{\rho}J_{-1}dx^--J_{0}d\rho,
\label{eq:APoin}
\end{align}
where $x^\pm=t\pm\phi$. The connections are
\begin{align}
\begin{aligned}
	e&=e^aJ_a=e^\rho J_\rho+e^tJ_t+e^\phi J_{\phi},\\
	\omega&=\omega^aJ_a=\omega^\rho J_\rho+\omega^tJ_t+\omega^\phi J_{\phi}
\end{aligned}
\end{align}
with
\begin{align}
\begin{aligned}
	&e^\rho=d\rho,\quad e^t=e^{\rho}dt,\quad e^\phi=e^{\rho}d\phi,\\
	&\omega^\rho=0,\quad \omega^t=e^{\rho}d\phi,\quad \omega^\phi=e^{\rho}dt.
\end{aligned}
\end{align}

Let us consider the boundary surface $Q$ defined by 
\ba
\phi=g(\rho).
\label{profgq}
\ea
The normal vector reads

\ba
n^\rho=-\frac{g'(\rho)}{\s{g'(\rho)^2+e^{-2\rho}}},\ \ \  n^t=0,\ \ \ 
n^\phi=\frac{e^{-\rho}}{\s{g'(\rho)^2+e^{-2\rho}}}.
\ea

Then the boundary condition (\ref{eq:NC1}) with the ansatz (\ref{profgq}) gives explicitly the equation for $a=t$, $a=\phi$ and $a=\rho$ as follows:
\ba
&& a=t:\ \ \ g'^3+2g'e^{-2\rho}+g''e^{-2\rho}+T(e^{-2\rho}+g'^2)^{3/2}=0,\no
&& a=\phi:\ \ \ g'+T\s{g'^2+e^{-2\rho}}=0,\no
&& a=\rho:\ \ \ g'+T\s{g'^2+e^{-2\rho}}=0.
\ea
These are simply solved as 
\ba
\phi=g(\rho)=\frac{T}{\s{1-T^2}}e^{-\rho}+\mbox{const.}
\label{profilegr}
\ea
which reproduces the known profile of boundary surface $Q$ in AdS/BCFT \cite{Takayanagi:2011zk,Fujita:2011fp}.

\subsection{Comment on Higher Spin Generalization}

It is intriguing to extend the above construction to the Chern-Simons description of higher spin gravity, whose CFT duals have been known \cite{Gaberdiel:2010pz,Gaberdiel:2012ku,Gaberdiel:2012uj} for a while.
We can construct a boundary action on $Q$ in higher spin gravity by replacing the SL($2,R$) indices  $a,b,c\ddd$ in $I_{GH}$ (\ref{eq:ICS2}), in addition to $I_{EH}$ (\ref{eq:ICS1}), with those for SL($n,R$) with $n=3,4,5,\ddd$, maintaining the higher spin gauge invariance (refer to appendix \ref{app:sl3} for SL($3,R)$). 

It is straightforward to find solutions which can also be embedded in SL($2,R$) theory like  (\ref{profilegr}), by solving the boundary conditions derived from this SL($n,R$) boundary action. However, it turns difficult to find genuine new solutions, which cannot be embedded in SL($2,R$) theory. We believe this is because we did not take into account higher spin charges in the boundary action.  We expect we can construct non-trivial solutions peculiar to higher spin gravity if we further add new terms  to  (\ref{eq:ICS2}) which will make solutions with higher spin charge consistent. This may be regarded as a generalization of the tension term in  (\ref{eq:ICS2}) for the SL($2,R$) theory. We would like to leave this as an important future problem.


\section{HEE in AdS/BCFT from Wilson Lines}

In this section, we extend the calculations of holographic entanglement entropy in higher spin gravity, 
using Wilson lines 
pioneered in  \cite{Ammon:2013hba,deBoer:2013vca} to our AdS/BCFT setups. The basic idea is that
a Wilson line describes a propagation of a massive particle. In AdS/CFT, such a massive particle is dual to 
a primary operator in a CFT. If we consider a twist operator which appears in the replica computation of 
entanglement entropy, as a primary operator, then we can read off the holographic entanglement entropy 
from the expectation value of Wilson line in the higher spin gravity. In our AdS/BCFT, we need to carefully treat the presence of the end of the world brane $Q$, on which the Wilson line ends. 
We will start our argument with the CS gravity with the gauge group SL($2,R$) and finally we extend it to higher spin gravity with the group
SL($3,R$). For explicit  calculations in this section, we choose the Poincare AdS$_3$ (\ref{AdS3m})
as a basic example of background. However, it should be straightforward to analyze more general examples
in our framework.

Let us first introduce the Wilson line operator \cite{Ammon:2013hba,deBoer:2013vca,Castro:2014mza}
\begin{align}
	W_R(C)=\langle U_f| P\exp\bigg(\int_CA\bigg)  P\exp\bigg(\int_C\bar{A}\bigg)  |U_i\rangle,
\label{eq:Wilson1}
\end{align}
where $R$ is a representation of the gauge group $G$, and $C$ is an open path from an initial state $|U_i\rangle$ to  a final state $|U_f\rangle$. Also $P$ means the path ordering. According to \cite{Ammon:2013hba}, the evaluation of Wilson line for a curve $C$ is also given by
\begin{align}
	W_R(C)=\int[DU][DP]e^{-I(U,P)_C}~,
\label{eq:Wilson2}
\end{align}
where $U(s)$ is the world line field on $s\in[s_i,s_f]$ and $P$ is a canonical momentum conjugate to $U$. Here $I(U,P)_C$ is the auxiliary system which lives on the Wilson line, and the path integral corresponds to the trace over $R$ in (\ref{eq:Wilson1}). In the saddle point approximation, this path integral gives the bulk geodesic length $D_{i,f}$ from $|U_i\rangle$ to $|U_f\rangle$ as
\begin{align}
	\log W_{R}(C)\propto D_{i,f}.
\end{align}
 
The Wilson line (\ref{eq:Wilson2}) depends on only the end point of path $C$ and the representation $R$. Indeed the former gives the length, while the latter contains the character of a massive particle like mass and spin. Thus the bulk conical singularity by the backreaction of Wilson lines is controlled by $R$. The highest weight of $R$ is labeled by the quadratic Casimir $c_2$ and indeed we may evaluate the entanglement entropy as
\begin{align}
	S_{EE}&=-\log\left(W_R(C)\right)=\sqrt{2c_2}D_{i,f}
\label{eq:WilsonEE}
\end{align}
with fixing $\sqrt{2c_2}\rightarrow c/6$.

\subsection{Geodesics between Bulk Two Points}
\label{sec:geodesic}

Firstly we give a simple example in Poincare AdS with no additional boundary $Q$. In the case of $G=SL(2,R)\times SL(2,R)$, the evaluation of Wilson line for a curve $C$ is given by the action \cite{Ammon:2013hba}:
\begin{align}
	I(U,P)_C=\int_Cds \left(\text{Tr}[PU^{-1}D_sU]+\lambda(s)(\text{Tr}[P^2]-c_2)\right),
\label{eq:AS2}
\end{align}
where $U$ and $P$ live in the group manifold $SL(2,R)$ and the Lie algebra $sl(2,R)$, and $D_sU=dU/ds+A_sU-U\bar{A}_s$ with $A_s\equiv A_\mu dx^\mu/ds$. Since the equation of motion for this action leads to
\begin{align}
	U^{-1}D_sU+2\lambda P=0,\quad \frac{dP}{ds}=0,\quad \text{Tr}[P^2]=c_2,
\end{align}
the on-shell action is evaluated as
\begin{align}
	I(U,P)_C|_{on-shell}=-2c_2\int_C ds \lambda(s).
\end{align}
In the paper \cite{Ammon:2013hba}, using this action, it was shown that the Wilson line which connects two points on the AdS boundary $\rho=\rho_\infty\to \infty$, correctly reproduces the geodesic length when we impose the Dirichlet boundary condition at the end points. Therefore the holographic entanglement entropy can be computed via a Wilson line.
Below in this subsection, as a first step, we will generalize this calculation to the case where the end points of Wilson line are both situated at the bulk AdS.

To begin with, let us consider 
possibilities of boundary conditions we can impose at the end points of Wilson line. 
If the curve has an end point, we need to impose
\begin{align}
\delta I(U,P)_C=	[PU^{-1}\delta U]=0,
\label{eq:BCUP}
\end{align}
at the end point in order to make the variation of the total action $\delta I(U,P)_C$ vanishes. This leads to either the Dirichlet boundary condition $\delta U=0$ or the Neumann boundary condition $P=0$.

For the Poincare solution of the Chern--Simons gravity we can write it in the following pure gauge form (\ref{eq:APoin}):
\begin{align}
	A=LdL^{-1},\quad \bar{A}=R^{-1}dR,
\end{align}
where $L$ and $R$ are given by
\begin{align}
	L=P\exp\left(-\int_{(0,0,0)}^{(t,\phi,\rho)}A\right),\quad R=P\exp\left(-\int^{(0,0,0)}_{(t,\phi,\rho)}\bar{A}\right).
\end{align}
Let us start with the trivial solution\footnote{Here ``trivial solution'' is the solution for $A=\bar{A}=0$ and ``actual solution'' is not. In \cite{Ammon:2013hba}, they also call this ``trivial solution'' as ``nothingness solution''.}
\begin{align}
	U_0(s)=u_0e^{-2\alpha(s)P_0},
\end{align}
where $\lambda=d\alpha/ds$, and indeed the actual solution and momentum are found by the gauge transformation as
\begin{align}
\begin{aligned}
	U(s)&=L(s)\cdot U_0\cdot R(s),\quad P(s)=R(s)^{-1}\cdot P_0 \cdot R(s).
\end{aligned}
\end{align}
Eliminating $u_0$ for $U(s_i)$ and $U(s_f)$, we obtain
\begin{align}
\begin{aligned}
	e^{-2(\alpha(s_f)-\alpha(s_i))P_0}=(R(s_i)U^{-1}(s_i)L(s_i))(R(s_f)U^{-1}(s_f)L(s_f))^{-1}.
\end{aligned}
\label{eq:defM}
\end{align}

We now impose the Dirichlet boundary condition $\delta U=0$ at both end points.  
In particular, we choose
\begin{align}
	U(s_i)=U(s_f)=1,
\end{align}
at both end points. Then we find (refer to (\ref{uip}))
\begin{align}
\begin{aligned}
	&e^{-2(\alpha(s_f)-\alpha(s_i))P_0}\\
	&\quad=(R(s_i)L(s_i))(R(s_f)L(s_f))^{-1}\\
	&\quad=\left(
	\begin{array}{cc}
	e^{\rho_f-\rho_i}+x_i^-(x_f^+-x_i^+)e^{\rho_i+\rho_f} & x_f^-e^{\rho_f-\rho_i}-x_i^-e^{\rho_i-\rho_f}+x_i^-x_f^-(x_f^+-x_i^+)e^{\rho_i+\rho_f}\\
      	(x_i^+-x_f^+)e^{\rho_i+\rho_f} & e^{\rho_i-\rho_f}+x_f^-(x_i^+-x_f^+)e^{\rho_i+\rho_f} \\
   	\end{array}
  	\right)
\end{aligned}
\label{eq:Mex1}
\end{align}
with the useful relation
\begin{align}
\begin{aligned}
	&e^{\alpha J_{\pm1}}={\bf 1}+\alpha J_{\pm1},\\
	&e^{\rho J_0}=\cosh\frac{\rho}{2}\cdot{\bf 1}+2\sinh\frac{\rho}{2}\cdot J_0.
\end{aligned}
\end{align}
By taking the trace, the LHS of (\ref{eq:Mex1}) becomes
\begin{align}
	\text{Tr}\big[ e^{-2(\alpha(s_f)-\alpha(s_i))P_0} \big]=2\cosh\big(-\sqrt{2c_2}(\alpha(s_f)-\alpha(s_i))\big)
\end{align}
and the RHS is
\begin{align}
	2\cosh(\rho_f-\rho_i)-(x_f^+-x_i^+)(x_f^--x_i^-)e^{\rho_i+\rho_f}=\frac{z_i^2+z_f^2-(x_f^+-x_i^+)(x_f^--x_i^-)}{z_iz_f}
\end{align}
with $z=e^{-\rho}$. Thus we obtain
\begin{align}
	\cosh\big(-\sqrt{2c_2}(\alpha(s_f)-\alpha(s_i))\big)=\frac{z_i^2+z_f^2-(x_f^+-x_i^+)(x_f^--x_i^-)}{2z_iz_f}.
\label{eq:geodesic1}
\end{align}
This manifestly shows that the value of $\sqrt{2c_2}|\alpha(s_f)-\alpha(s_i)|$ coincides with the geodesic length (refer also to the appendix \ref{apGS} for a brief summary of geodesic length in AdS$_3$).

If we choose $s_i$ and $s_f$ to be at the AdS boundary such that $\rho_{i}=\rho_f=\rho_\infty\to \infty$, we can evaluate the entanglement entropy in AdS$_3$ such as
\begin{align}
	S_{EE}=I(U,P)_C|_{on-shell}=\frac{c}{3}\log\left( \frac{\s{(\Delta\phi)^2-(\Delta t)^2}}{\epsilon} \right)
\label{conee}
\end{align}
where we set $\sqrt{2c_2}=c/6$, $\epsilon=e^{-\rho_\infty}$, $\Delta t=t_f-t_i$, and $\Delta\phi=\phi_f-\phi_i$.

\subsection{Holographic Entanglement Entropy in AdS/BCFT from Chern-Simons Gravity}
\label{sec:HEE}

Now we would like to get back to our original problem. We would like to formulate the calculation of holographic entanglement entropy in the presence of the boundary surface $Q$ within our CS gravity formalism. The AdS/BCFT prescription argues that the holographic entanglement entropy
 when the subsystem A is a half line, is computed from the length of the geodesic which connects a boundary point and a point on $Q$ as we reviewed in section \ref{sec:heeb}. Even for generic choices of subsystem $A$ in two dimensional BCFTs, the holographic entanglement entropy is basically given by combining the result in this half line case  and the 
result for a connected geodesic  (\ref{conee}).

\begin{figure}
  \centering
  \includegraphics[bb=0 0 960 540,width=8cm]{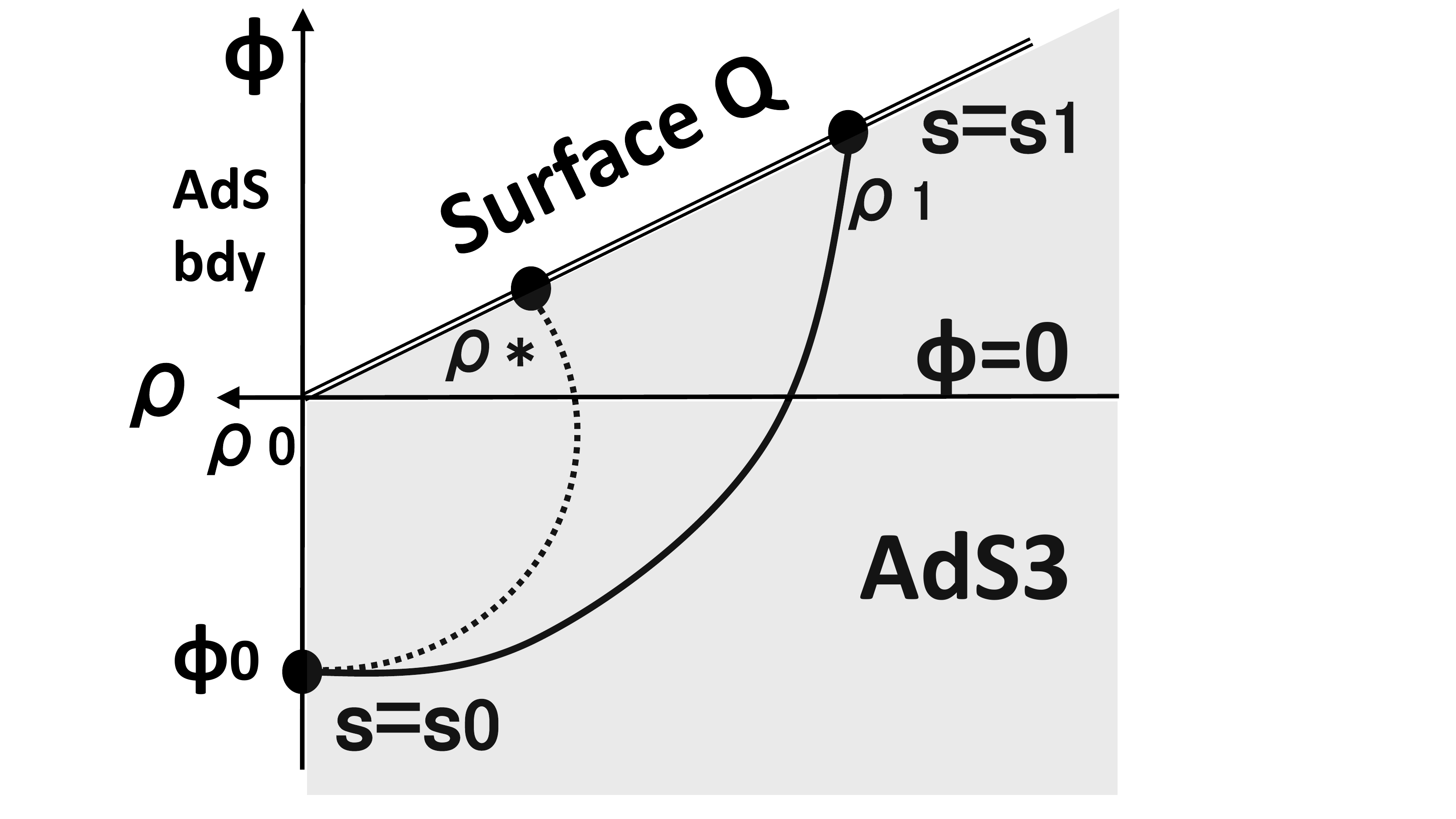}
  \caption{A sketch of a Wilson line which computes the entanglement entropy in AdS/BCFT. The colored region is the physical spacetime for the AdS/BCFT. The dotted curve describes the minimal length one. 
We omit the time direction. In CS gravity description, for any choices of a point $Q$ the Wilson line takes
the same expectation value once we impose our Neumann-Dirichlet boundary condition.}
\label{bsetupTfig}
\end{figure}

We express the initial point at $s_i=s_0$ and the final point at $s_f=s_1$ in AdS$_3$ (\ref{AdS3m}) 
by the coordinates
$(t_0,\rho_0,\phi_0)$ and $(t_1,\rho_1,\phi_1)$, respectively.
 We take $(t_0,\rho_0,\phi_0)$ to be at the AdS boundary
i.e. $\rho_0\to \infty$ and take $(t_1,\rho_1,\phi_1)$ to be on the surface $Q$ i.e. $\phi_1=\frac{T}{\s{1-T^2}}e^{-\rho_1}$ in (\ref{profilegr}). Refer to Fig. \ref{bsetupTfig} for a sketch.

Remember that we can write $U(s)$ as follows
\ba
U(s)=P\exp\left(-\int^{(t(s),\rho(s),\phi(s))}_{(0,0,0)} A_\mu dx^\mu \right)\cdot u_0 e^{-2\ap(s)P_0} \cdot 
P\exp\left(-\int^{(0,0,0)}_{(t(s),\rho(s),\phi(s))} \bar{A}_\mu dx^{\mu}\right),\no
\label{unitss}
\ea
and this allows us to write the variation of the end point $(t(s),\rho(s),\phi(s))$ as
\ba
\delta U=-(A_\mu U-U\bar{A}_\mu)dx^\mu.  \label{variua}
\ea

Let us first count the number of integration constants in the solution (\ref{unitss}) for the interval $s_0\leq s\leq s_1$, which should all be fixed after imposing a complete set of boundary conditions. There are totally six integration constants i.e. the $SL(2,R)$ matrix $u_0$ (three degrees of freedom) and the traceless matrix $P_0$ 
with the constraint  $\mbox{Tr}[P^2]=c_2$ (two degrees of freedom), and finally the value of $\ap(s_1)$. 
Note that we simply set 
\ba
\ap(s_0)=0,  \label{conder}
\ea
because this can be absorbed into $u_0$.\footnote{We do not care the intermediate values 
$\ap(s)$ for $s_0<s<s_1$ as they do not affect the on-shell values of the action. They can be gauged away.}

First we impose the Dirichlet boundary condition 
\ba
U(s_0)=1,  \label{condiy}
\ea
as we did before for the end points on the AdS boundary. 
Together with (\ref{conder}), this boundary condition (\ref{condiy}) fixes
\ba
u_0=e^{J_1x^+_0}e^{2 J_0\rho_0}e^{J_{-1}x^-_0},  \label{ucho}
\ea
via the relation (\ref{unitss}).
 This determines three out of six constants. 

\begin{figure}
  \centering
  \includegraphics[bb=0 0 960 540,width=8cm]{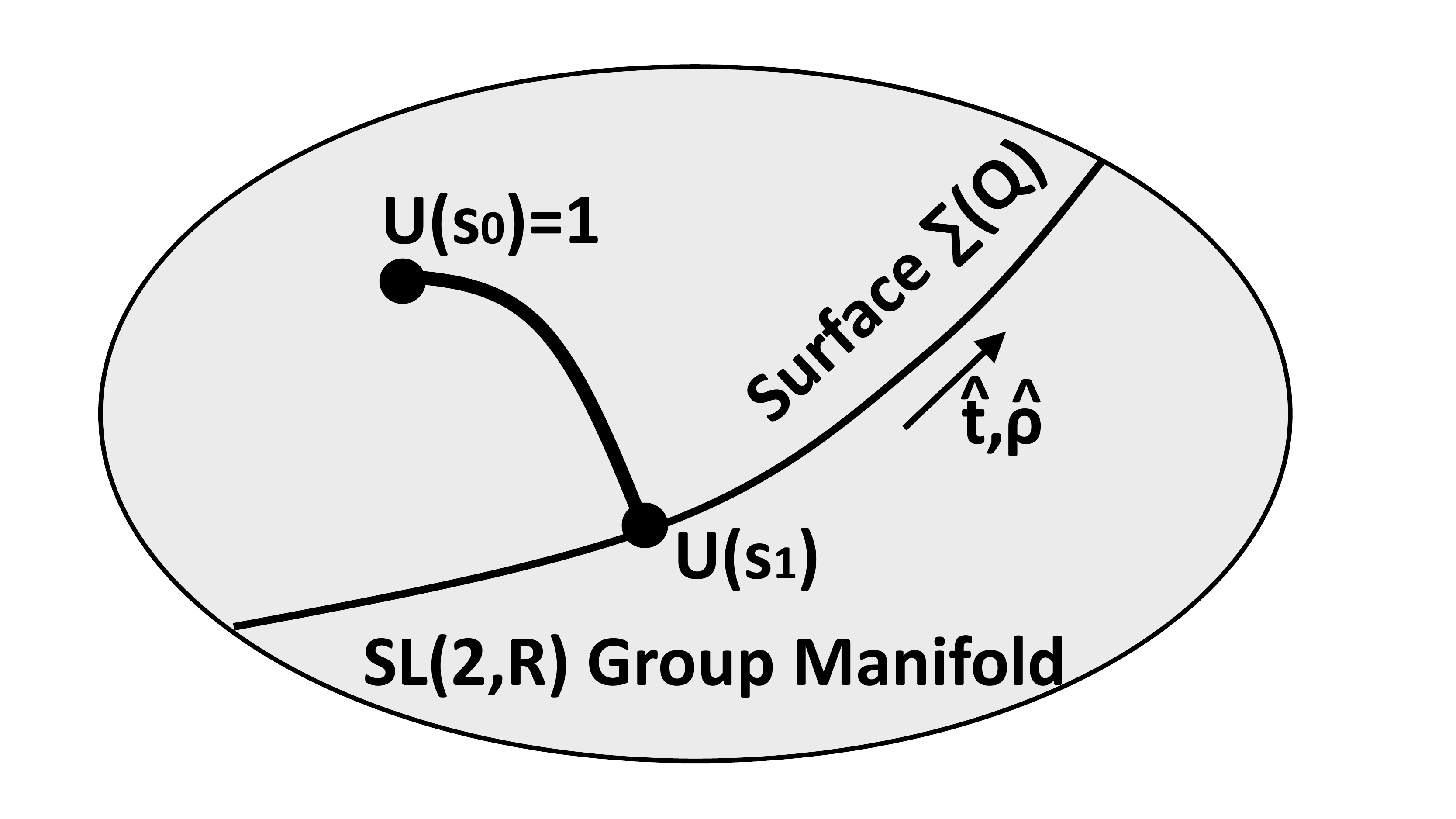}
  \caption{A sketch of computation of holographic entanglement entropy from the view point of 
SL(2,R) group manifold. A trajectory of $U(s)$ in the SL(2,R) group manifold is depicted as a thick curve with dots at its ends. We impose the initial condition $U(s_0)=1$ and 
the Dirichlet-Neumann boundary condition such that $U(s_1)$ is on $\Sigma_Q$, which is a codimension one 
subspace of SL(2,R).}
\label{openfig}
\end{figure}

We argue the remaining three boundary conditions are given by a mixed Dirichlet-Neumann boundary condition 
at $s=s_1$, whose detail will be described as follows. First, let us introduce the parameters of SL(2,R) manifold 
 $(\hat{t},\hat{\rho},\hat{\phi})$ and write the constant matrix part of $U(s)$:
\ba
u_0 e^{-\ap(s_1)P_0}\equiv e^{J_1 \hat{x}^+}e^{2J_0\hat{\rho}}e^{J_{-1}\hat{x}_-}. \label{relxp}
\ea
In other words, we can parameterize the matrix $U(s_1)$ we are interested in as follows
\ba
U(s_1)= e^{-J_0 \rho_1}e^{-J_1x^+_1}\cdot e^{J_1 \hat{x}^+}e^{2J_0\hat{\rho}}e^{J_{-1}\hat{x}^-}\cdot
e^{-J_{-1}x^-_1}e^{-J_0\rho_1}.\label{unitsqss}
\ea
For each value of $(t_1,\rho_1,\phi_1)$,  our Dirichlet boundary condition is to require that the point 
$(\hat{t},\hat{\rho},\hat{\phi})$ is included in the surface $Q$ given by (\ref{profilegr}) i.e. 
\ba
\hat{\phi}=\frac{T}{\s{1-T^2}}e^{-\hat{\rho}}.  \label{qsura}
\ea
Let us define the codimension one subspace $\Sigma(Q)$ of SL($2,R$) as (refer to Fig.\ref{openfig})
\ba
\Sigma(Q)=\{ e^{J_1 \hat{x}^+}e^{2J_0\hat{\rho}}e^{J_{-1}\hat{x}_-}\in \mbox{SL($2,R$)} \ |\ (\hat{t},\hat{\rho},\hat{\phi})\in Q\}.  
\ea
This is a group counterpart of the surface $Q$ in the AdS/BCFT. Then our Dirichlet boundary condition is equivalently written as 
\ba
e^{J_1 x^+_1}e^{J_0\rho_1}\cdot U(s_1)\cdot e^{J_0\rho_1}e^{J_{-1}x^-_1}=e^{J_1 \hat{x}^+}e^{2J_0\hat{\rho}}e^{J_{-1}\hat{x}_-} \in \Sigma(Q).  \label{drfg}
\ea
This is a natural condition by considering the behavior of Wilson lines when we embed the boundary surface $Q$ in the target space namely AdS$_3$=SL($2,R$).

Next we impose the Neumann condition in the direction parallel to the ``D-brane" $\Sigma(Q)$. This can be done by remembering  that the variation of $\mbox{Tr}[PU^{-1}\delta U]$ should vanish (\ref{eq:BCUP}), where 
the variation $\delta U$ is given by the variation of (\ref{unitsqss}) by shifting the values of 
$(\hat{t},\hat{\rho},\hat{\phi})$ with the constraint (\ref{qsura}) satisfied. This is equivalent to the variation in 
 (\ref{variua})  along the surface Q. Thus the Neumann boundary condition is found as  
\ba
v^\mu \mbox{Tr}[P (U^{-1}A_\mu U-\bar{A}_\mu)]\bigr|_Q=0,  \label{eqvmu}
\ea
where $v$ is a vector parallel to the surface $Q$ and thus (\ref{eqvmu}) is decomposed into two equations
on $Q$ defined by the coordinate $(t_1,\rho_1,\phi_1=\frac{T}{\s{1-T^2}}e^{-\rho_1})$:
\ba
&& \mbox{Tr}[P(U^{-1}A_tU-\bar{A}_t)]=0,\label{atbc} \\
&& \mbox{Tr}[P(U^{-1}A_\rho U-\bar{A}_\rho)]-\frac{T}{\s{1-T^2}}e^{-\rho_1} \cdot \mbox{Tr}[P(U^{-1}A_\phi U-\bar{A}_\phi)]=0.\label{arbc}
\ea
These conditions guarantee that the on-shell action (i.e. the expectation value of Wilson line)  is independent from the values of the end point $(t_1,\rho_1)$ on $Q$. Indeed these conditions can be obtained from 
(\ref{unitss}) by taking variation of $(x^+(s_1),x^{-}(s_1),\rho(s_1))$ along the surface $Q$, with $\alpha(s_1)$ and $P_0$ kept fixed.

Now we are in a position to solve the above two Neumann boundary conditions (\ref{atbc}), (\ref{arbc})  and 
one Dirichlet boundary condition (\ref{drfg}), which totally provide three constraints we wanted.
The two Neumann equations (\ref{atbc}) and (\ref{arbc}) fix $P_0$ as follows
\ba
P_0=\s{\frac{c_2}{2}}\frac{1}{\s{e^{-2\rho_0}+\phi_0^2}}\left(
 \begin{array}{cc}
-t_0 &  e^{-2\rho_0}+\phi_0^2-t_0^2 \\
1 & t_0\\
    \end{array}
  \right), \label{vapz}
\ea 
where this is independent of $t_1$ and $\rho_1$.

Finally we would like to solve the Dirichlet boundary condition, which should fix $\ap(s_1)$.
Before that let us note that interestingly,
we can show by using (\ref{ucho}) and (\ref{vapz}) that (\ref{relxp}) is equivalent to
\ba
\hat{t}=t_0,\ \ \ \ \phi_0^2+e^{-2\rho_0}=\hat{\phi}^2+e^{-2\hat{\rho}},
\ea
which shows  that $(\hat{t},\hat{\rho},\hat{\phi})$ is on the geodesic which connects 
two boundary points $(t_0,\rho_0,\phi_0)$ and $(t_0,\rho_0,-\phi_0)$. 
Then, our Dirichlet boundary condition, which requires that 
the point  $(\hat{t},\hat{\rho},\hat{\phi})$ is on the surface $Q$ i.e. 
$\hat{\phi}=\frac{T}{\s{1-T^2}}e^{-\hat{\rho}}$,  fixes the value of $(\hat{t},\hat{\rho},\hat{\phi})$ to be 
$(t_*,\rho_*,\phi_*)$ given by
\ba
t_*=t_0, \ \ \ \ \ e^{-\rho_*}=\s{(1-T^2)(e^{-2\rho_0}+\phi_0^2)},\ \ \ \ \ \phi_*=T\s{e^{-2\rho_0}+\phi_0^2}.  \label{dotp}
\ea
This point $(t_*,\rho_*,\phi_*)$ actually coincides with the special point on $Q$ which has the minimal distance to the boundary point $(t_0,\rho_0,\phi_0)$ among all points on $Q$ on the time slice $t=t_0$
(refer to Fig.\ref{bsetupTfig}). This is the essential reason why this prescription of calculation the 
holographic entanglement entropy via the Wilson line, matches with the standard one in Einstein gravity 
reviewed in section \ref{sec:heeb}.

After imposing all these conditions, the matrix $U(s_1)$ is completely fixed as follows: 
\ba
U(s_1)= e^{-J_0 \rho_1}e^{-J_1x^+_1}\cdot e^{J_1 x^+_*}e^{2J_0\rho_*}e^{J_{-1}x^-_*}\cdot
e^{-J_{-1}x^-_1}e^{-J_0\rho_1}.\label{unitsss}
\ea
It is important to notice that when $(t_1,\rho_1,\phi_1)=(t_*,\rho_*,\phi_*)$, we find $U(s_1)=1$. This corresponds to the point on the surface $Q$ which minimize the length of the geodesic.

At the same time, the equation (\ref{relxp}) fixes the value of $\ap(s_1)$ to be
\ba
e^{\s{2c_2}\ap(s_1)}=(\s{1+\phi_0^2e^{2\rho_0}}-\phi_0e^{\rho_0})\cdot \s{\frac{1+T}{1-T}},
\ea
which leads to the correct geodesic length and the HEE in the limit $\rho_0\to \infty$ (assume $\phi_0<0$):
\ba
S_A=\frac{c}{6}\log \left(2|\phi_0|e^{\rho_0}\right)+\frac{c}{6}\log\s{\frac{1+T}{1-T}}.
\label{eq:EEQ}
\ea
This perfectly reproduces the known expression (\ref{bcftee}) by setting $\ep=e^{-\rho_0}$.
The boundary entropy is given by 
\ba
S_{bdy}=\frac{c}{6}\log\s{\frac{1+T}{1-T}},
\ea
which agrees with the result in  \cite{Takayanagi:2011zk,Fujita:2011fp}.

In this way, we obtain the correct prescription of holographic entanglement entropy for AdS/BCFT  via Wilson lines in the CS gravity, such that the results agree with the earlier prescription \cite{Takayanagi:2011zk,Fujita:2011fp} in Einstein gravity.
In the usual AdS/BCFT for Einstein gravity, it is given by the length of geodesic which departs from the AdS boundary and ends on the surface $Q$ and we minimize the length by changing the location of endpoint on $Q$. In our new calculation in CS gravity, we consider a trajectory of $U(s)$ from the boundary point $s=s_0$ to the surface $\Sigma_{Q}$ in the SL(2,R) group manifold. Note that the result does not depend on the end point of the Wilson line, as opposed to the standard holographic entanglement entropy for Einstein gravity.  
In our CS calculation, the end of the world brane $Q$ looks like  a "D-brane" for a particle moving in the SL($2,R$) group manifold as in Fig.\ref{openfig}.

\subsection{Comments on Extension to Higher Spin Gravity}
It is intriguing to extend the above prescription of computing holographic entanglement entropy to 
that for higher spin gravity.
In the case of $G=SL(3,R)\times SL(3,R)$, the evaluation of Wilson line is also given by the action 
\cite{Ammon:2013hba}
\begin{align}
	I(U,P)_C=\int_Cds \left(\text{Tr}[PU^{-1}D_sU]+\lambda_2(s)(\text{Tr}[P^2]-c_2)+\lambda_3(s)(\text{Tr}[P^3]-c_3)\right),
\label{eq:AS3}
\end{align}
where  we define $\text{Tr}[P^2]\equiv P^aP^b\delta_{ab}$ and $\text{Tr}[P^3]\equiv P^aP^bP^ct_{abc}$, see appendix \ref{app:sl3} for details. To evaluate (\ref{eq:AS3}), we use the equation of motion
\begin{align}
	D_sP=0,\quad U^{-1}D_sU+2\lambda_2(s)P+3\lambda_3(s)P\times P=0,\quad \text{Tr}[P^2]=c_2,\quad \text{Tr}[P^3]=c_3
\end{align}
with the definition
\begin{align}
	P\times P=t_{abc}T^aP^bP^c.
\end{align}
The on-sell action becomes
\begin{align}
\begin{aligned}
	I(U,P)_C|_{on-shell}&=-\int_Cds(2c_2\lambda_2(s)+3c_3\lambda_3(s)).
\end{aligned}
\end{align}
Note here that we demand $c_3=0$ for the purpose of computing entanglement entropy. If we consider the case of $c_3\neq0$, we should give the different interpretation for expectation value of Wilson lines instead of the entanglement entropy in the dual CFTs\footnote{It means that the entanglement entropy does not carry the higher spin charge, see \cite{Ammon:2013hba,Castro:2014mza} for extensive discussion.}.

We also start with the trivial solution
\begin{align}
	U_0=u_0e^{-2\alpha_2(s)P_0-3\alpha_3(s)(P_0\times P_0)}.
\end{align}
As in section \ref{sec:geodesic}, we find that the evaluated on-shell action is given by
\begin{align}
\begin{aligned}
	I(U,P)_C|_{on-shell}&=-2c_2\Delta\alpha_2-3c_3\Delta\alpha_3,
\end{aligned}
\label{eq:I3}
\end{align}
and obtain the relation
\begin{align}
\begin{aligned}
	e^{-2\Delta\alpha_2P_0-3\Delta\alpha_3(P_0\times P_0)}&=(R(s_i)U^{-1}(s_i)L(s_i))(R(s_f)U^{-1}(s_f)L(s_f))^{-1},
\end{aligned}
\label{eq:defM3}
\end{align}
with $\Delta\alpha_N=\alpha_N(s_f)-\alpha_N(s_i)$. Using the above two equations, we find 
\begin{align}
\begin{aligned}
	S_{EE}=I(U,P)_C|_{on-shell}=\text{Tr}\left[P_0 \log\Big((R(s_i)U^{-1}(s_i)L(s_i))(R(s_f)U^{-1}(s_f)L(s_f))^{-1}\Big)  \right],
\end{aligned}
\label{eq:General}
\end{align}
which is the general expression for $G=SL(N,R)\times SL(N,R)$ since this is independent of $G$, see also \cite{Castro:2014mza}.

Let us next consider the case with the boundary surface $Q$. According to the section \ref{sec:HEE}, the boundary condition for $U(s)$ is given by (using (\ref{unitsss}))
\begin{align}
	U(s_0)=1,\quad U(s_1)=e^{-J_0 \rho_1}e^{-J_1x^+_1}\cdot e^{J_1 x^+_*}e^{2J_0\rho_*}e^{J_{-1}x^-_*}\cdot
e^{-J_{-1}x^-_1}e^{-J_0\rho_1},
\end{align}
where $(\rho_*,x^{\pm}_*)$ are given by the same values (\ref{dotp}). Note that this again follows from the Dirichlet-Neumann boundary condition (\ref{drfg}), which is embedded into 
SL($2,R$) in an obvious way.
Thus we obtain the following result for AdS/BCFT in higher spin gravity:
\begin{align}
\begin{aligned}
	S_{EE}=\text{Tr}\left[P_0 \log\Big((R(s_0)L(s_0))(R(s_*)L(s_*))^{-1}\Big)  \right]
\end{aligned}
\end{align}
which reproduce the proper geodesic length and reduces to (\ref{eq:EEQ}) for $N=2$.

Since the precise boundary action in CS formalism is not available at present, we do not know 
what profiles of the end of the world brane are allowed in the presence of higher spin fields. Therefore the above analysis is limited to the case where we can embed the solution to the subgroup SL($2,R$). 
Nevertheless, we may expect to obtain the boundary condition we need to impose for Wilson lines 
by a straightforward generalization of the ``D-brane condition"  (\ref{drfg}) from SL($2,R$) to SL($n,R$).
We would like to leave a full analysis of holographic entanglement entropy for higher spin gravity for 
an important future problem.

	
\section{Conclusion and Discussion}

In this paper, we presented a formulation of three dimensional AdS/BCFT in terms of  Chern-Simons (CS) gravity. In the first half part, we gave the correct boundary action of CS gravity which precisely reproduces the AdS/BCFT  in Einstein gravity. For this we introduce a dynamical vector on the end of the world brane $Q$ and impose boundary conditions by varying the total action including the boundary terms. In the later half part, we presented the prescription of computing the holographic entanglement entropy in our CS gauge theoretic description of AdS/BCFT. By employing the Wilson line calculations, the holographic entanglement entropy $S_A$ for a subsystem $A$ can be obtained from a Wilson line which connects the AdS boundary point and a point on the end of the world brane $Q$. The former coincides with the point $\de A$ as usual and the latter can be an arbitrary point on $Q$. The most non-trivial ingredient is the boundary condition at the point on $Q$ for the Wilson line calculation. We argue that this boundary condition is given by a mixed Neumann-Dirichlet boundary condition, where the end of the world brane $Q$ looks like a ``D-brane" in the SL($2,R$) group manifold, if we think the Wilson line is analogous to an open string. We also confirmed that our prescription in CS gravity correctly reproduces the know result in BCFTs.  Moreover, we gave an extension of our prescription of holographic entanglement entropy to that for higher spin gravity in a simple example.

We expect our results in this paper provide an important first step when we try to fully generalize the AdS/BCFT in higher spin gravity, where the boundary surface $Q$ can have higher spin charges. For example, our boundary action 
(\ref{eq:ICS2}) and our Neumann-Dirichlet boundary condition: (\ref{atbc}), (\ref{arbc})  and (\ref{drfg}),
at a point on $Q$ for the Wilson line computation, have a straightforward generalization from the SL($2,R$) to  SL($n,R$) group. Even though we might expect some additional ingredients, these will provide useful clues
to formulate the AdS/BCFT for higher spin gravity. A non-trivial test of correctness of holographic entanglement entropy in higher spin gravity will be the strong subadditivity (SSA). This is because in the 
Wilson line formalism, the derivation of SSA looks difficult as there is no minimization procedure involved, 
while in the standard holographic entanglement entropy, the proof of SSA follows directly from the minimal surface property \cite{Headrick:2007km,Wall:2012uf}. 
We would like to come back to these important problems in our future publication.

\subsection*{Acknowledgements}

We would like to thank Yasuaki Hikida for various discussions.
TT is supported by the Simons Foundation through the ``It from Qubit'' collaboration.  
TT is supported by Inamori Research Institute for Science and 
World Premier International Research Center Initiative (WPI Initiative) 
from the Japan Ministry of Education, Culture, Sports, Science and Technology (MEXT). 
TT is supported by JSPS Grant-in-Aid for Scientific Research (A) No.~16H02182 and 
by JSPS Grant-in-Aid for Challenging Research (Exploratory) 18K18766.
TU is supported by the Grant-in-Aid for JSPS Research Fellow, No.19J11212.

\appendix

\section{Lie Algebra}
\subsection{$sl(2)$ Conventions}
\label{app:sl2}

We denote the generators of $sl(2,R)$ as $J_a$, which satisfy
\begin{align}
[J_a,J_b]=\epsilon_{abc}J^c.
\end{align}
In the case of the fundamental representation of $sl(2)$, the generators are written as
\begin{align}
 J_1=\left(
 \begin{array}{cc}
  0 & 0 \\
  -1 & 0 \\
 \end{array} 
\right),\quad J_{0}=\left(
 \begin{array}{cc}
  1/2 & 0 \\
  0 & -1/2 \\
 \end{array} 
\right),\quad J_{-1}=\left(
 \begin{array}{cc}
  0 & 1 \\
  0 & 0 \\
 \end{array} 
\right).
\label{eq:sl2base}
\end{align}
The metric is defined as
\begin{align}
	\eta_{ab}=2\text{Tr}[J_aJ_b].
\end{align}

In our calculations of this paper,  the following identity is useful:
\ba
e^{J_{-1}\ap}e^{-2\rho J_0}e^{-J_1\beta}e^{2\eta J_0}e^{-J_{-1}\gamma}
 = \left(
    \begin{array}{cc}
      e^{\eta-\rho}+e^{\eta+\rho}\ap\beta & e^{-\eta+\rho}\ap-e^{\eta-\rho}\gamma-e^{\eta+\rho}\ap\beta\gamma  \\
      e^{\eta+\rho}\beta & e^{-\eta+\rho}-e^{\eta+\rho}\beta\gamma \\
    \end{array}
  \right).  \label{uip}
\ea

\subsection{$sl(3)$ Conventions}
\label{app:sl3}

We also denote the generators of $sl(3)$ as $T_a=\{ L_i,W_m \}$ with $i=-1,0,1$ and $m=-2,\dots,2$, which satisfies
\begin{align}
\begin{aligned}
	&[L_i,L_j]=(i-j)L_{i+j},\\
	&[L_i,W_m]=(2i-m)W_{i+m},\\
	&[W_m,W_n]=-\frac{1}{3}(m-n)(2m^2+2n^2-mn-8).
\end{aligned}
\end{align}

In the case of fundamental representation of $sl(3)$, the generators are written as
\begin{align}
 \begin{aligned}
  L_1&=\left(
  \begin{array}{ccc}
   0 & 0 & 0 \\
   1 & 0 & 0 \\
   0 & 1 & 0 \\
  \end{array} 
  \right),\quad L_0=\left(
  \begin{array}{ccc}
   1 & 0 & 0 \\
   0 & 0 & 0 \\
   0 & 0 & -1 \\
  \end{array} 
  \right),\quad L_{-1}=\left(
  \begin{array}{ccc}
   0 & -2 & 0 \\
   0 & 0 & -2 \\
   0 & 0 & 0 \\
  \end{array} 
  \right),\\
  W_2&=\left(
  \begin{array}{ccc}
   0 & 0 & 0 \\
   0 & 0 & 0 \\
   2 & 0 & 0 \\
  \end{array} 
  \right),\quad W_1=\left(
  \begin{array}{ccc}
   0 & 0 & 0 \\
   1 & 0 & 0 \\
   0 & -1 & 0 \\
  \end{array} 
  \right),\quad W_{0}=\frac{2}{3}\left(
  \begin{array}{ccc}
   1 & 0 & 0 \\
   0 & -2 & 0 \\
   0 & 0 & 1 \\
  \end{array} 
  \right),\\
  W_{-1}&=\left(
  \begin{array}{ccc}
   0 & -2 & 0 \\
   0 & 0 & 2 \\
   0 & 0 & 0 \\
  \end{array} 
  \right),\quad W_{-2}=\left(
  \begin{array}{ccc}
   0 & 0 & 8 \\
   0 & 0 & 0 \\
   0 & 0 & 0 \\
  \end{array} 
  \right).
 \end{aligned}
\label{eq:sl3base}
\end{align}
The symmetric tensors are defined as
\begin{align}
	\delta_{ab}=\frac{1}{2}\text{Tr}[T_aT_b],\quad h_{abc}=\text{Tr}[T_{(a}T_bT_{c)}].
\end{align}


\section{Chern--Simons Formulation of Einstein Gravity}
\label{app:CSform}

As (\ref{eq:IEH}) and (\ref{eq:IGH}), the three dimensional gravity on $\mathcal{M}$ is written by the metric formulation:
\begin{align}
\begin{aligned}
	I_{tot}&=I_{EH}+I_{GH},\\
	&=\frac{1}{16\pi G}\int_{\mathcal{M}} d^3x \sqrt{-g}(R-2\Lambda)+\frac{1}{8\pi G}\int_{\partial \mathcal{M}} d^2x \sqrt{-g}K.
\end{aligned}
\end{align}
Here the Gibbons--Hawking term $I_{GH}$ needs to be added in order to make the variation principle sensible since the variation of $I_{EH}$ contains $\delta\partial_\rho g_{ij}$ and this term spoils a variational principle on $\partial\mathcal{M}$. Thus we find
\begin{align}
	\delta I_{tot}=\frac{1}{16\pi G}\int_M\sqrt{-g}(K_{ij}-K g_{ij})\delta g^{ij},
\end{align}
and it leads to the two types of conditions
\begin{align}
\begin{aligned}
	& \mbox{Dirichlet}:~\delta g^{ij}=0, \\
	& \mbox{Neumann}:~K_{ij}-K g_{ij}=0.
\end{aligned}
\end{align}
Note here that we chose the following coordinate system
\begin{align}
	ds^2=g_{\mu\nu}dx^\mu dx^\nu=d\rho^2+g_{ij}dx^idx^j.
\end{align}

\subsection{Vielbein Formalism}

We may translate from the metric formalism to the vielbein formalism with the basic variables, which consists of vielbeins  $e^a_\mu$ and spin connections $\omega^a_{\mu b}$. The metric $g_{\mu\nu}$ is written in terms of vielbeins as
\begin{align}
	g_{\mu\nu}=e^a_{\mu}e^b_{\nu}\eta_{ab},
\end{align}
and covariant derivative of $V^a=V^\mu e^a_\mu$ and $V^a_b=V^\mu_\nu e^a_\mu e^\nu_b$ would be
\begin{align}
\begin{aligned}
	&\nabla_\mu V^a=\partial_\mu V^a+\omega^a_{\mu b}V^b,\\
	&\nabla_\mu V^a_b=\partial_\mu V^a_b+\omega^a_{\mu c}V^c_b-\omega^c_{\mu b}V^a_c.
\end{aligned}
\end{align}
Then, the Riemann tensor can be written in terms of the spin connections as
\begin{align}
	R=d\omega+\omega\wedge\omega,
\end{align}
or alternatively $R^a_{\mu\nu b}=\partial_\mu\omega^a_{\nu b}-\partial_\nu\omega^{a}_{\mu b}+[\omega_\mu,\omega_\nu]^a_b$. 

With above setups, the Einstein-Hilbert action is written as
\begin{align}
\begin{aligned}
	I_{EH}&=\frac{1}{16\pi G}\int_\mathcal{M} \epsilon_{abc}e^a\wedge \left(R^{bc} -\frac{\Lambda}{3} e^b\wedge e^c\right)\\
	&=\frac{1}{16\pi G}\int_\mathcal{M}\left[ e^a\wedge \left(2d\omega_a+\epsilon_{abc}\omega^b\wedge\omega^c\right)-\frac{\Lambda}{3}\epsilon_{abc}e^a\wedge e^b\wedge e^c\right],
\end{aligned}
\label{eq:IV}
\end{align}
where we use the relation
\begin{align}
	d^3x&=\frac{1}{3!}\epsilon_{\mu\nu\sigma}dx^\mu\wedge dx^\nu\wedge  dx^\sigma,\\
	\sqrt{-g} &=|e|=-\frac{1}{3!}\epsilon_{abc}\epsilon^{\mu\nu\sigma}e^a_\mu e^b_\nu e^c_\sigma,\\
	R^{ab}&=d\omega^{ab}+\omega^a_c\wedge\omega^{cb},\\
	\omega^a&=\frac{1}{2}\epsilon^{abc}\omega_{bc}.
\end{align}
The the Gibbons--Hawking term is also rewritten
\begin{align}
\begin{aligned}
	I_{GH}=-\frac{1}{8\pi G}\int_{\partial\mathcal{M}}P_{ab}e^a\wedge\omega^b-\frac{1}{8\pi G}\int_{\partial\mathcal{M}}\epsilon_{abc}e^a\wedge n^bdn^c.
\end{aligned}
\end{align}
Here we use the following relations
\begin{align}
	K&=e^i_a(\partial_in^a+\epsilon^a_{~cb}\omega_{i}^{c}n^b),\\
	d^2x&=\frac{1}{2}\epsilon_{ij}dx^i\wedge dx^j,\\
	\sqrt{-g}&=\frac{1}{2}\epsilon_{abc}\epsilon^{ij}e^a_i e^b_jn^c,
\end{align}
and find
\begin{align}
\begin{aligned}
	d^2x\sqrt{-g}K=-\epsilon_{abc}e^a\wedge\left(n^b\epsilon^c_{~de}\omega^dn^e +n^bdn^c \right).
\end{aligned}
\end{align}
Thus the variation of $I_{tot}$ leads to the condition (\ref{eq:NC1}), (\ref{eq:NC2}) and (\ref{eq:NC3}) with $T=0$.

\subsection{Chern--Simons Formalism}
The Chern--Simons action with  the boundary given in (\ref{eq:ICS1}) and (\ref{eq:ICS2}) is obtained from (\ref{eq:IV}) as
\begin{align}
\begin{aligned}
	I_{EH}=&\frac{k}{4\pi }\int_\mathcal{M}\text{Tr}\left[ A\wedge dA +\frac{2}{3}A\wedge A\wedge A\right] - \frac{k}{4\pi }\int_\mathcal{M}\text{Tr}\left[ \bar{A}\wedge d\bar{A} +\frac{2}{3}\bar{A}\wedge \bar{A}\wedge \bar{A}\right]\\
	&\qquad+ \frac{k}{4\pi }\int_{\partial\mathcal{M}}\text{Tr}\left[A\wedge\bar{A}\right]
\end{aligned}
\end{align}
with $k=1/(4G)$, $A=A^aJ_a$ and
\begin{align}
	\text{Tr}\left[ J_aJ_b \right]=\frac{1}{2}\delta_{ab},\quad \text{Tr}\left[ J_aJ_b J_c\right]=\frac{1}{4}\epsilon_{abc}.
\end{align}
Here we use the relation
\begin{align}
	&A^a\wedge dA_a-\bar{A}^a\wedge d\bar{A}_a=\frac{2}{\ell}\left( \omega^a\wedge de_a +e^a\wedge d\omega_a\right)\\
	&d(A^a\wedge\bar{A}_a)=dA^a\wedge \bar{A}_a-A^a\wedge d\bar{A}_a=\frac{2}{\ell}\left(-\omega^a\wedge de_a+e^a\wedge d\omega_a\right).
\end{align}
Thus the first term of (\ref{eq:IV}) is
\begin{align}
	e^a\wedge d\omega_a=\frac{1}{4}\left( A^a\wedge dA_a-\bar{A}^a\wedge d\bar{A}_a+ d(A^a\wedge\bar{A}_a)\right),
\end{align}
and the second and third ones are
\begin{align}
	\epsilon_{abc}e^a\wedge\left(\omega^b\wedge\omega^c+\frac{1}{3}e^b\wedge e^c  \right)=\frac{\epsilon_{abc}}{6}\left(A^a\wedge A^b\wedge A^c-\bar{A}^a\wedge\bar{A}^b\wedge \bar{A}^c\right).
\end{align}
The Gibbons--Hawking term is also rewritten
\begin{align}
\begin{aligned}
	I_{GH}=-\frac{k}{2\pi }\int_{\partial\mathcal{M}}\text{Tr}\left[A\wedge\bar{A}\right]+\frac{1}{8\pi G}\int_{\partial\mathcal{M}}e^a\wedge n^b\left(n_a\omega_b-\epsilon_{abc} dn^c\right),
\end{aligned}
\end{align}
where we use the relation
\begin{align}
\begin{aligned}
	P_{ab}e^a\wedge\omega^b=e^a\wedge\omega_a-e^an_a\wedge\omega_bn^b.
\end{aligned}
\end{align}

Thus the variation of  $I_{tot}$ leads to the condition
\begin{align}
\begin{aligned}
	& \mbox{Dirichlet}:~\delta(A^a-\bar{A}^a)=0,
\end{aligned}
\end{align}
and (\ref{eq:NC1}) with $T=0$.


\section{Geodesic Length}\label{apGS}

Consider the metric of global AdS$_{3}$ 
\ba
ds^2=-\cosh^2\rho d\tau^2+d\rho^2+\sinh^2\rho d\theta^2,
\ea
and  Poincare AdS$_{3}$:
\ba
ds^2=\frac{dz^2-dt^2+d\phi^2}{z^2}.
\ea
The geodesic length $D$ between two bulk points $(\rho_1,\tau_1,\theta_1)$ and $(\rho_2,\tau_2,\theta_2)$
in the global AdS$_3$ 
is given by 
\ba
\cosh D=\cosh\rho_1\cosh\rho_2\cos(\tau_1-\tau_2)-\sinh\rho_1\sinh\rho_2\cos(\theta_1-\theta_2).
\ea
In the same way the geodesic distance $D=\cosh\sigma$ between two bulk points $(z_1,\phi_1,t_1)$ and 
$(z_2,\phi_2,t_2)$ in the Poincare AdS$_3$ is found as
\ba
\cosh D=\frac{z_1^2+z_2^2+(\phi_1-\phi_2)^2-(t_1-t_2)^2}{2z_1z_2}.
\ea

\bibliographystyle{JHEP}
\bibliography{CSBCFT}

\end{document}